# Si CMOS Platform for Quantum Information Processing

L. Hutin[1], R. Maurand[2], D. Kotekar-Patil[2], A. Corna[2], H. Bohuslavskyi[1,2],
X. Jehl[2], S. Barraud[1], S. De Franceschi[2], M. Sanquer[2], M. Vinet[1]

[1] CEA, LETI, Minatec Campus, F-38054 Grenoble, France    [2] CEA, INAC-PHELIQS, F-38054 Grenoble, France
e-mail: louis.hutin@cea.fr ; silvano.defranceschi@cea.fr

**Abstract**

We report the first quantum bit (qubit) device implemented on a foundry-compatible Si CMOS platform. The device, fabricated using SOI NanoWire MOSFET technology, is in essence a compact two-gate pFET. The qubit is encoded in the spin degree of freedom of a hole Quantum Dot (QD) defined by one of the Gates. Coherent spin manipulation is performed by means of an RF E-Field signal applied to the Gate itself. By demonstrating qubit functionality in a conventional transistor-like layout and process flow, this result bears relevance for the future up-scaling of qubit architectures, including the opportunity of their co-integration with "classical" Si CMOS control circuitry.

## Introduction

Owing to the quantum superposition and entanglement principles, an N-qubit state is characterized by $2^N$ complex coefficients corresponding to the normalized probabilities for each possible arrangement of basis states (**Figure 1**). This built-in parallelism in information treatment, if harnessed by proper algorithms [1], holds great promise for a variety of applications such as secure data exchange, database search, machine learning, and simulation of quantum processes. Electron spin qubits in coupled semiconductor Quantum Dots [2] increasingly emerge as a promising alternative to the leading superconducting-based solid-state approach. Notably, recent work in nuclear spin-free isotopically pure $^{28}$Si has exhibited remarkably long quantum coherence times [3],[4]; benefitting from a fundamental advantage over hitherto prevalent III-V materials [5]. This has spurred a sudden interest in Si-based spin qubits in the fundamental research community. At a practical level though, a major remaining challenge thus far was the prototyping of a truly foundry-compatible quantum information building block, compact enough to consider circuits controlling hundreds of qubits and beyond. The following results represent a significant step in this direction.

## Gate-Defined Quantum Dots

Prior to manipulating its spin, a first requirement is the ability to confine an elementary charge in a potential well separated from carriers reservoirs by tunnel barriers. A single dopant atom in the Si lattice can provide an ultimately abrupt potential well, but such an approach requires the development of deterministic doping techniques [6]. Ours consists in using the Gates of MOSFETs to create Quantum Dots (QDs) and leveraging the Coulomb Blockade (CB) effect. **Figure 2** shows how NanoWire (NW) FETs with Wrap-Around Gate and wide spacers on thin undoped SOI are particularly suitable for observing well resolved Coulomb diamonds. Our process fabrication flow and cross sections **Figure 3** illustrate the practical implementation; a thermal oxidation was performed after mesa patterning in order to further reduce the NW diameter. A 7nm SiO$_2$ thickness was left before Gate stack deposition in order to keep the high-κ interface away from the dot in an effort to limit noise and charge fluctuations. The obtained Wrap-around Gate configuration compensates for the EOT increase and contributes to improving the electrostatic control of the QD by the Gate. Wide spacers keep stray dopants away from the dot, reduce the lateral $C_S$ and $C_D$, and provide sufficiently large resistance to achieve proper confinement. Note that the backside substrate can also be used as an additional Gate to chisel the potential profile along the channel. The most aggressively scaled devices (L=10nm, NW Ø=3.4nm) exhibited Coulomb oscillations even at room temperature [7]. However a mere metallic island-like CB behavior is not sufficient for our purpose since the ability to split the "spin-up" and "spin-down" energy levels requires a discrete energy spectrum in the dot. In the range of manufacturable L×W dimensions, such quantization can be resolved at very low temperatures, typically below 1K. **Figure 4** shows the Coulomb diamonds obtained by plotting the conductivity vs. ($V_{GS}$, $V_{DS}$) in our SOI NWpFETs at T=100mK (L=25nm, $W_{top}$=15nm). This type of stability diagram maps the number of elementary charges stored in the QD. After single charge confinement, the next step is to evidence a successful spin manipulation, which requires the implementation of a spin-charge conversion scheme.

## Spin-charge conversion: Pauli Spin Blockade in Coupled QDs

A geometry of two adjacent Gates over a common NW leads to form two coupled dots in series (**Figure 5**). In this configuration, the Source-Drain current may be sensitive to spin orientation due to the Pauli Spin Blockade (PSB) effect [8]. According to the Pauli Exclusion Principle, two identical fermions (*eg* holes) cannot coexist in the same quantum state. Hence, if both QDs are in the same spin state, no charge transfer can occur between them. The signature of two coupled dots in series is well-known as a honeycomb network with double-triangles **Figure 6**, further details can be found in [8]. A manifestation of PSB can be observed at the boundary between charge states "(1,1)" and "(0,2)" as the disappearing base of the double-triangle. Therefore in the established range of ($V_{G1}$, $V_{G2}$, $V_{DS}$) PSB biasing conditions, an induced spin rotation in one of the two dots will be detected as a current increase. Note that quantum coupling between QDs is one of the more difficult issues to tackle. In the future, long-range coupling solutions may be key enablers towards circuit upscaling. A 65nm Gate pitch seems to be sufficient to guarantee nearest neighbor interactions at the considered temperatures. This was achieved in this work via e-beam lithography, though these dimensions are within reach of conventional optical lithography with multiple patterning [9].

## Coherent Spin Control by RF Electric Field

Lifting the spin degeneracy (**Figure 7**) can be done by applying a static magnetic field B, resulting in a separation of the "spin-up" and "spin-down" states by the Zeeman energy $E_Z=|g|\mu_B B$, where g is a dimensionless quantity related to the gyromagnetic ratio of the particle, and $\mu_B$ is the Bohr magneton. Spin resonance is triggered if the energy of an electromagnetic excitation wave h.f matches exactly $E_Z$. The straightforward way to manipulate the spin is to induce electron or hole spin resonance via RF magnetic field [3],[4]. However this implies i/ coping with non-local effects of the magnetic field bleeding out to influence neighboring devices, ii/ additional strip lines preferably made of a superconducting material to limit Joule effect thermal dissipation. Alternately, under some conditions such as the existence of spin-orbit coupling or the possibility to perform g-tensor modulation resonance (g-TMR), an E-field RF excitation alone may also lead to spin transitions (cf. electrons in InAs [10], holes in InSb [11]). In Si, spin-orbit coupling is present in the Valence Band [12]. It was also recently shown that the criteria were met for the "g-TMR" phenomenon to occur for holes in our pFET devices [13]. Electric-field manipulation can thus be implemented by applying an RF signal directly to the MOSFET Gate defining the tested QD. Indeed, a line of higher Drain-to-Source conductivity does appear when plotting the PSB current against the strength of the static magnetic field B for various frequencies, verifying the Zeeman energy equation (**Figure 8**). It is the first observation of an Electrically-Driven Spin Resonance (EDSR) for a hole QD in Si. A potential interest of manipulating holes is their alleged immunity to nuclear spin-mediated decoherence which may occur via hyperfine coupling in natural Si [14]. The EDSR signal itself is not a proof that coherent control is achieved, *i.e.* that the qubit state can be placed arbitrarily on the Bloch sphere **Figure 1**. The spin rotation angle is function of the RF burst duration. Using the scheme described in **Figure 9**, inter-dot charge movement can be suppressed by sitting in CB conditions during a spin manipulation, before lifting it and reading the PSB current. Said current versus τ$_{burst}$ exhibits the so-called Rabi oscillations **Figure 10**, for which the maxima correspond to a π (mod 2π) spin rotation. Characteristic of Rabi oscillations, the period and thus the spin manipulation speed is proportional to $P^{1/2}$. In summary, this shows that we were able to initialize the spin of a single charge isolated in a quantum dot, control the weighted superposition of its "spin-down" and "spin-up" states, couple it to a near neighbor and read it out.

## Perspectives and Conclusion

We successfully demonstrated the first electrically-driven hole spin qubits in Si using SOI NanoWire CMOS technology, and leveraging the Field-Effect for both charge confinement and spin manipulation. We described one way to detect a change in spin orientation via DC transport. More advanced techniques enabling high-fidelity single-shot readout are currently under investigation, such as dispersive readout/RF reflectometry [15],[16]. Regarding the prospect of future possible co-integration of qubits with classical control electronics, we have recently reported a first demonstration of a ring oscillator capable of operating down to 1K [17]. Though we are still witnessing the very early stages of Si-based electron/hole spin-based quantum information, these results suggest that the experience and know-how accumulated while chasing Moore's law might prove unexpectedly useful in the materialization of profoundly new computational paradigms in the decades to come.

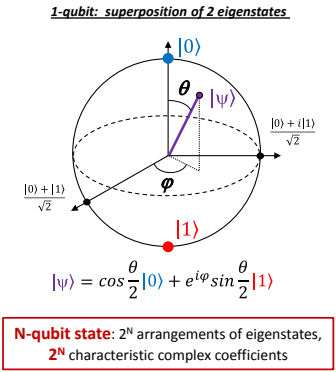

Fig.1: Bloch sphere representation of the quantum state space. A qubit state can be described by a linear combination of eigenstates e.g. "spin-up" and "spin-down".

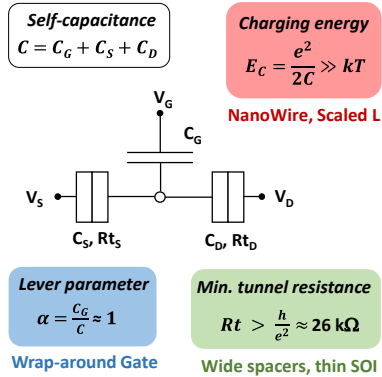

Fig.2: A Gate-controlled Quantum Dot flanked by tunnel junctions exhibits Coulomb Blockade (CB) operation when $E_c \gg kT$, which means it is possible to control the exact number of charges in the dot.

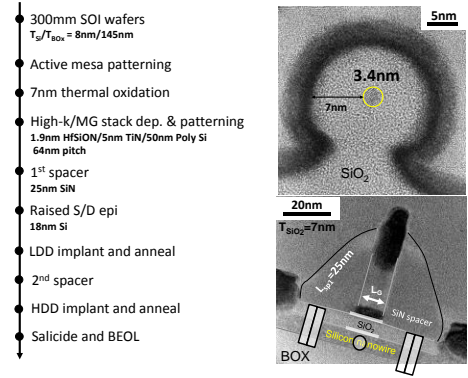

Fig.3: Simplified process flow and TEM cross-sections along the Wrap-around Gate (top) and along the channel (bottom) of SOI NanoWire pFETs. Wide spacers are primarily used for proper Gate-defined dot confinement.

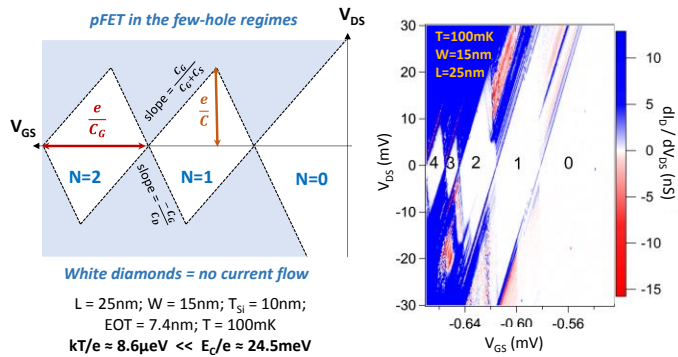

Fig.5: Implementation of spin-charge conversion scheme through Pauli Spin Blockade (PSB). a) Top view SEM of a device with two gates in series (L=30nm, W=30nm, inter-Gate spacing $S_{gg}$=35nm). The wide spacers protect the inter-Gate spacing from dopant implantation. b) cross-sectional sketch along the channel showing the Valence Band profile with the sub-spacer tunnel barriers and the quantization of states in the Gate-defined dots. (c) Principle of Pauli Spin Blockade conditioning the transition from the (1,1) to (0,2) charge states.

Fig. 4: Left – principle of the conductivity mapping vs. ($V_{GS}$, $V_{DS}$) in a pFET, exhibiting Coulomb diamonds of zero conductivity. The N=0 diamond is wider due to the sub-threshold regime. A better resolution is achieved when $\alpha\approx1$, $E_C$ is large and T is low. Right – experimental data from a single SOI NWpFET (L=25nm; W=15nm) measured at T=100mK.

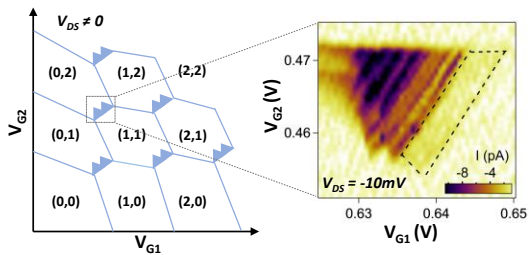

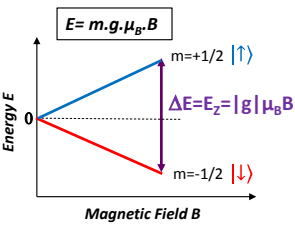

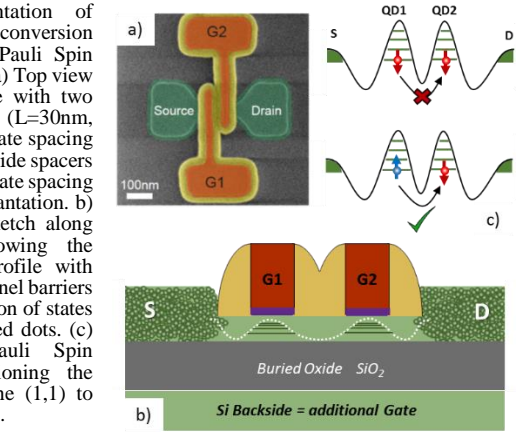

Fig.6: Left – Principle of the conductivity mapping vs ($V_{G1}$, $V_{G2}$). A characteristic honeycomb network is obtained with "double-triangles", the bases of which correspond to ground state transport. Right – experimental measurement at the "(1,1)", "(0,2)" transition showing evidence of Pauli Spin Blockade, i.e. a disappearing base. The ³He measurement setup base temperature is 15mK in the following.

Fig.7: Principle of Zeeman splitting for spin-½ particles: lifting of the spin degeneracy under a static magnetic field. The energy separation is the $|g|\mu_B B$ product.

Fig.8: Measured Electrically Driven Spin Resonance (EDSR) signals. Spin transitions can occur if the energy of an EM excitation matches the Zeeman energy. Here, the Pauli Spin Blockade is lifted using only an E-field excitation on the G1 Gate. Plotting the current versus static magnetic field strength for various E-field frequencies yields spikes located along a line verifying the equation $hf=E_Z=|g|\mu_B B$.

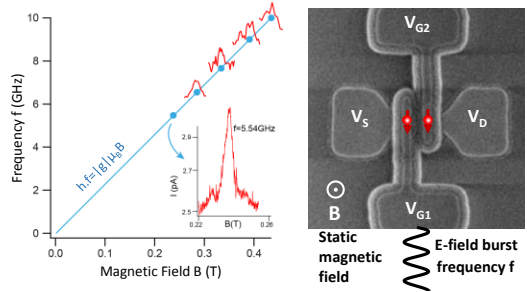

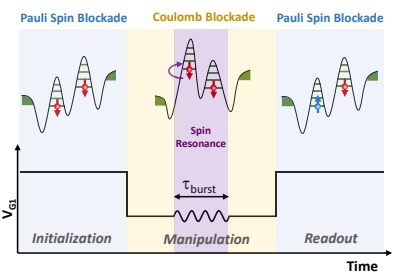

Fig.9: Principle of the protocol for resolving Rabi oscillations. The QD potentials are detuned in CB conditions during spin manipulation to prevent current from flowing before the readout step.

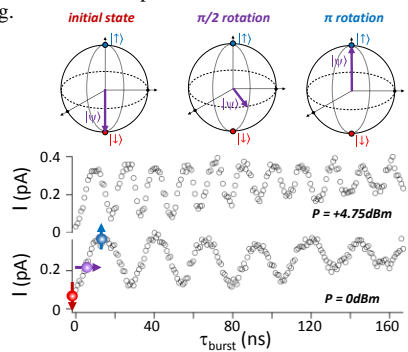

Fig.10: Rabi oscillations of measured Source-Drain current vs. $\tau_{burst}$ (see Fig. 9). The period varies with $P^{1/2}$. This demonstrates coherent control of the hole spin orientation by the Gate.

**Acknowledgment** – The authors acknowledge financial support from the EU under Projects SiSPIN (No. 323841) and SiAM (No. 610637).